\documentclass[runningheads]{llncs}
\usepackage{graphicx}
\usepackage{amsmath}
\usepackage{subfigure}
\usepackage{multirow}
\usepackage{booktabs} 
\usepackage{bm}
\usepackage{float}

\begin{document}
\title{Polar-Net: A Clinical-Friendly Model for Alzheimer's Disease Detection in OCTA Images}

\titlerunning{Polar-Net: A Clinical-Friendly Model for Alzheimer's Disease Detection in OCTA Images}

\author{
Shouyue Liu\inst{1,2} \and
Jinkui Hao\inst{1} \and
Yanwu Xu\inst{3,4*} \and
Huazhu Fu\inst{5} \and
Xinyu Guo\inst{1} \and
Jiang Liu\inst{6} \and
Yalin Zheng\inst{7} \and
Yonghuai Liu\inst{8} \and
Jiong Zhang\inst{1} \and
Yitian Zhao\inst{1*}
}

\authorrunning{Shouyue Liu et al.}
%
\institute{Cixi Institute of Biomedical Engineering, Ningbo Institute of Materials Technology and Engineering, Chinese Academy of Sciences \email{yitian.zhao@nimte.ac.cn}
\and Cixi Biomedical Research Institute, Wenzhou Medical University
\and School of Future Technology, South China University of Technology, Guangzhou
\and Pazhou Lab,Guangzhou
\email{ywxu@ieee.org}
\and Institute of High-Performance Computing, Agency for Science, Technology and Research, Singapore
\and Department of Computer Science, Southern University of Science and Technology
\and Department of Eye and Vision Science, University of Liverpool
\and Department of Computer Science, Edge Hill University
}

\maketitle

\begin{abstract}

Optical Coherence Tomography Angiography (OCTA) is a promising tool for detecting Alzheimer's disease (AD) by imaging the retinal microvasculature. Ophthalmologists commonly use region-based analysis, such as the ETDRS grid, to study OCTA image biomarkers and understand the correlation with AD. 
In this work, we propose a novel deep-learning framework called Polar-Net. Our approach involves mapping OCTA images from Cartesian coordinates to polar coordinates, which allows for the use of approximate sector convolution and enables the implementation of the ETDRS grid-based regional analysis method commonly used in clinical practice. Furthermore, Polar-Net incorporates clinical prior information of each sector region into the training process, which further enhances its performance. Additionally, our framework adapts to acquire the importance of the corresponding retinal region, which helps researchers and clinicians understand the model's decision-making process in detecting AD and assess its conformity to clinical observations.
Through evaluations on private and public datasets, we have demonstrated that Polar-Net outperforms existing state-of-the-art methods and provides more valuable pathological evidence for the association between retinal vascular changes and AD. In addition, we also show that the two innovative modules introduced in our framework have a significant impact on improving overall performance.

\end{abstract}

\section{Introduction}
Alzheimer's disease (AD) is a progressive and debilitating neurological disorder that affects millions of people worldwide. 
Although primary detection of AD can be achieved through a combination of cognitive function tests and neuroimaging techniques,  such as magnetic resonance imaging (MRI) and cerebrospinal fluid (CSF) analysis \cite{palmer2022cognitive}. However, these approaches suffer from being invasive, time-consuming, or expensive, hindering their use in routine clinical practice.
The convergence of tissue origin, structural characteristics, and functional mechanisms between the eyes and the brain has been previously reported~\cite{teja2017cerebral}. For example, patients with AD have significantly decreased blood vessel density in superficial parafoveal and choriocapillaris (CC) \cite{zhang2021choriocapillaris}.
To this end, the automated AD detection using fundus image has emerged as an active research field in the last two years\cite{ma2021rose,10.1007/978-3-031-16434-7_66,cheung2022deep}. Color fundus photography (CFP) has commonly used for AD studies, but the CFP has limitations in capturing the information of deep layer vessels. Optical coherence tomography angiography (OCTA) is an innovative non-invasive technology that generates high-resolution images of depth-resolved retinal microvasculature projections \cite{jeong2022determining}, including SVC, DVC, and CC.

\begin{figure}[!t]
\centering
\includegraphics[width=1\linewidth]{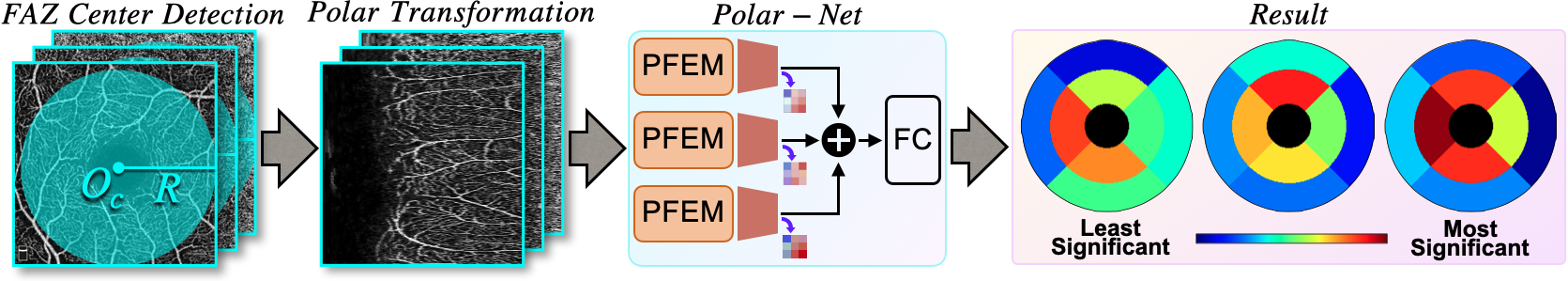}
\vskip -3pt
\caption{The proposed AD detection model utilizes the polar transformation (left two) inspired by ETDRS grids commonly used in ophthalmic analysis. This model allows for easy understanding by ophthalmologists, as the output of the Polar-Net can be interpreted through an intuitive color map illustration (right) that indicates different levels of significance. }
\label{figflowchart}
\vspace{-5pt}
\end{figure}

Studies on clinical biomarkers of OCTA images are mainly based on regional analysis, e.g., the early treatment of diabetic retinopathy study (ETDRS) grid, which divides a target area into 9 regions with three concentric circles and two orthogonal lines, as shown in the right three sub-figures in Fig.~\ref{figflowchart}.
The region-based analysis allows a more specific evaluation of retinal changes and their correlation with AD, which can provide a more nuanced understanding of the disease. %
Research following ETDRS and IE grid demonstrated the significance of many regions, e.g., in the three sub-regions of nasal-outer, superior-inner, and inferior-inner in inner vascular complexes, which present a substantial decrease in vascular area density and vascular length density for the AD participants \cite{xie2023deep}.

Over the past few years, deep-learning-based algorithms have achieved remarkable success in the analysis of medical images. As for AD detection, several methods use an integration of multiple modalities\cite{wisely2022convolutional,10.1007/978-3-031-16434-7_66}.
However, these methods rarely follow the clinical region-based analysis routine, which limits their ability to incorporate valuable clinical statistical findings and generate easily interpretable results.
To address the above issues, we proposed a novel deep-learning framework to take full advantage of clinical region-based analysis, for AD detection in OCTA images. To obtain a more accurate and interpretable result, we specifically designed an approximate sector convolution, based on the polar transformation and a multi-kernel feature extraction module. The main contributions of the paper can be summarised as follows:
\textbf{(1)} Based on the well-known clinically used ETDRS grids for retinal image analysis, we incorporate the regional importance prior in the training process through a weight matrix, so as to better understand the correlations between retinal structure alternations and AD.
\textbf{(2)}
We introduce an approximate sector convolution through polar transformation, to mimic the clinical region-based analysis, by mapping the OCTA image from the Cartesian system to the polar system, as shown in the left two sub-figures in Fig.~\ref{figflowchart}. 
\textbf{(3)} We further performed the explainability analysis on the well-trained model. The interpretable results showed consistency with the conclusions of the previous clinical studies, indicating that the proposed method can be a potential tool, to investigate the pathological evidence of the relationship between the fundus and AD.

\vspace{-10pt}
\section{Methodology}

Fig.~\ref{figflowchart} shows the flowchart of our AD detection method using SVC, DVC, and CC projections of OCTA as input. First, we utilize VAFF-Net \cite{hao2022retinal} to locate the center of the FAZ on SVC. We then transform the original images into the polar coordinates with the FAZ center as the origin. 
The transformed images are then fed into our Polar-Net, which produces the final detection result and the corresponding region importance matrix.
\vspace{-10pt}
\subsection{Polar coordinate transformation for OCTA image}

\begin{figure}[!t]
\centering
\includegraphics[width=0.8\linewidth]{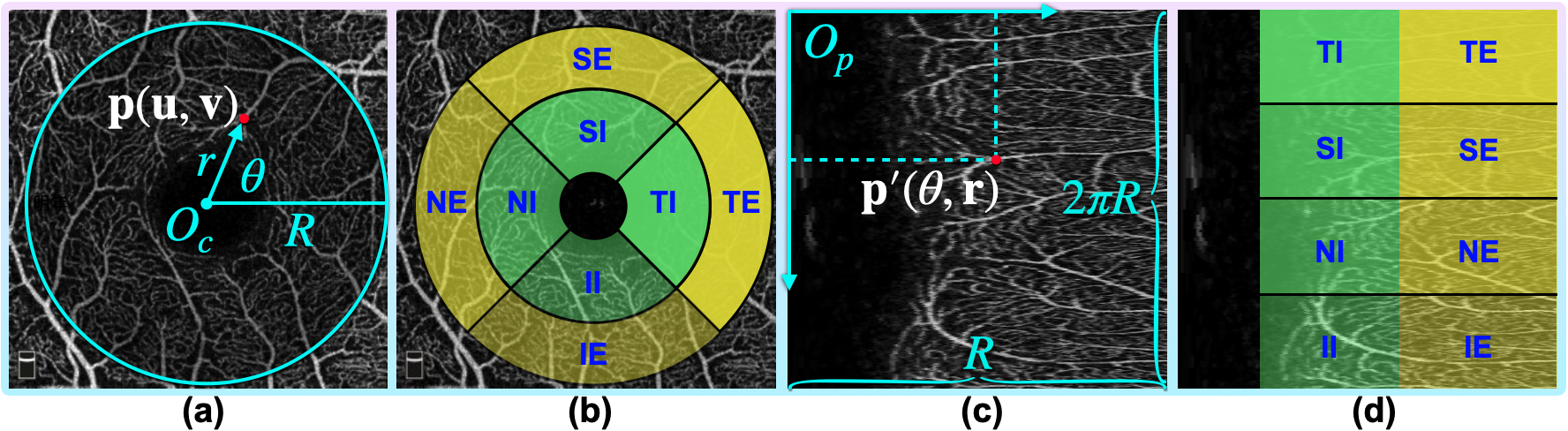}
\vskip -3pt
\caption{
Illustrations of the Cartesian coordinate system (a), an ETDRS grid on an OCTA projection (b), the polar coordinate system (c), and the mapping relationship after the transformation (d). Definition: temporal-inner (TI), temporal-external (TE), superior-inner (SI), superior-external (SE), nasal-inner (NI), nasal-external (NE), inferior-inner (II), and inferior-external (IE).}
\label{figpolartrans}
\vspace{-5pt}
\end{figure}

We introduce a method called polar transformation, to realize region-based analysis. As shown in Fig. \ref{figpolartrans}, the polar transformation converts the region of interest (blue circle) into a polar coordinate system Fig. \ref{figpolartrans}(c), with the center of the FAZ (according to the definition\cite{early1991early} of ETDRS), ${O_c(u_o, v_o)}$ as the origin. The original image is represented as points in the Cartesian system ${p(u, v)}$, and the corresponding points in the polar system are represented by ${p\prime(\theta,r)}$. 
The relationship between these two coordinate systems is given by the following equations:
\begin{equation}\small
\left\{\begin{array} { l } 
{ u = r \operatorname { cos } \theta } \\
{ v = r \operatorname { sin } \theta }
\end{array} \Leftrightarrow \left\{\begin{array}{l}
r=\sqrt{u^{2}+v^{2}} \\
\theta=\tan ^{-1} v / u
\end{array}\right.\right. \   .
\end{equation}
The width of the transformed image is equal to the distance ${R}$, the minimal length from the center ${O_c(u_o, v_o)}$ to the edge in the original image, and the height is ${2\pi R}$. Since the corners are cropped, the outermost pixels of the region of interest are kept in order to preserve the original information as much as possible, and the part near ${O_c}$ is filled by nearest neighbor interpolation. The polar transformation represents the original image in the polar coordinate system by pixel-wise mapping\cite{fu2018joint}, and has the following properties:

\noindent \textbf{1) Approximate sector-shaped convolution.} Convolution is widely used in convolutional neural networks (CNNs), where the shape of the convolution kernel is always rectangular. However, in the real world, many semantics are non-rectangular, such as circle and sector, which makes the adaptability of the receptive field in CNNs suboptimal. For the polar transformation, the mapping relationship is fixed, enabling us to approximate the sector convolution with a rectangular convolution kernel at a lower computational cost. The mapping relationship shown in Fig. \ref{figpolartrans}(b)(d) explains this, and for the sake of simplicity and clarity, we use ETDRS girds as an example. When we perform convolution with a rectangular kernel along the ${TI \rightarrow SI}$ direction on the transformed image, it is equivalent to performing convolution with a sector-shaped kernel counterclockwise around the FAZ center in the original image.

\noindent \textbf{2) Equivalent augmentation.} Applying data augmentation to the original image is the same as applying data augmentation in the polar system since the transformation is a pixel-wise mapping \cite{fu2018joint}. 
For instance, by changing the start angle and the transformation center ${O_c(u_o, v_o)}$, we can realize the drift cropping operation in the polar system. It is analogous to applying various cropping factors for data augmentation by changing the transformation radius ${R}$.

\begin{figure}[!t]
\centering
\includegraphics[width=1\linewidth]{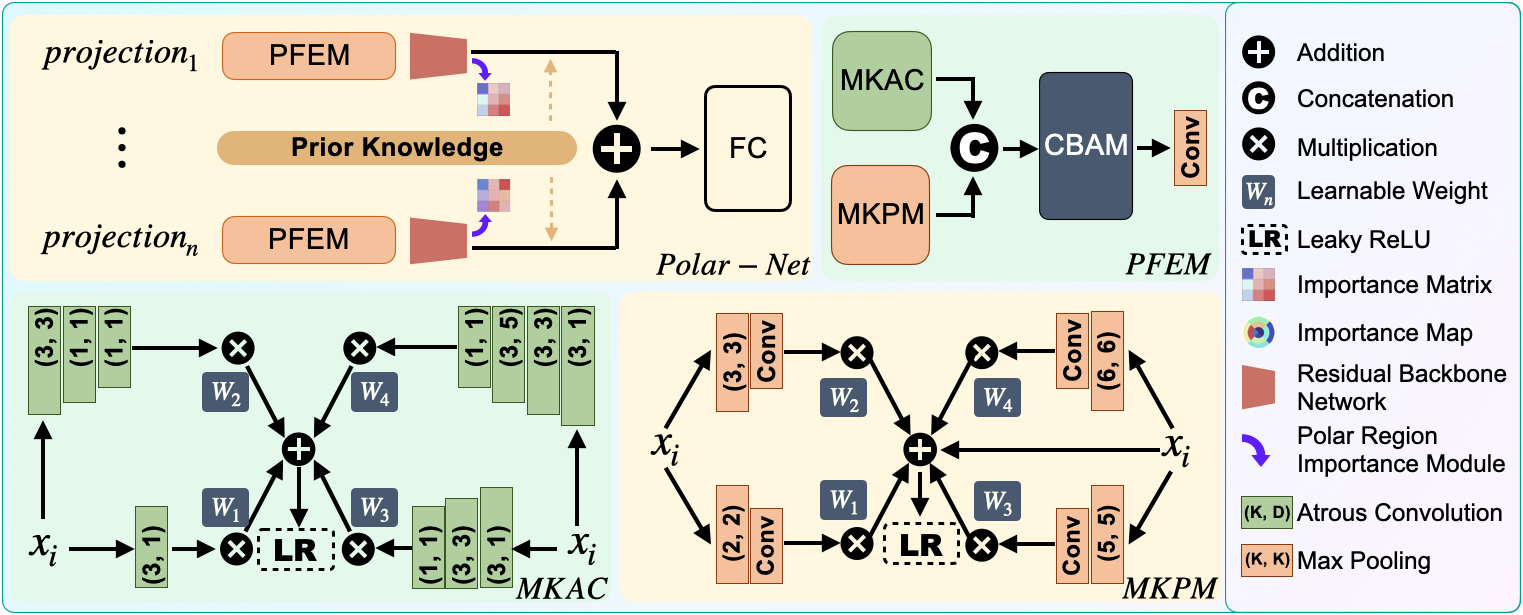}
\vskip -3pt
\caption{The details of Polar-Net. It contains multiple input branches, and each branch starts with a polar feature extractor module (PFEM) and ends with a residual network. PFEM consists of a multi-kernel atrous convolution module (MKAC), a multi-kernel pooling module (MKPM), and a convolutional block attention module (CBAM). Parameters ${K}$ and ${D}$ denote the kernel size and dilation respectively.}
\label{fignetdetail}
\vspace{-5pt}
\end{figure}

\subsection{Network architecture}
In the transformed image, we can extract features around the FAZ. Rectangular features at different scales correspond to different sectoral features in the retina. Therefore, it is critical to extract information across different sizes of the visual field. To this end, we design the Polar-Net. As shown in Fig \ref{fignetdetail}, it contains several branches, the number of which varies according to the number of projections. Each branch starts with a polar feature extractor module (PFEM) and ends with a residual network. To take full advantage of all the branches, middle fusion is used. To generate the region importance matrix, a polar region importance module (PRIM) is proposed, which follows the residual network. Furthermore, Polar-Net can receive a prior knowledge matrix to utilize clinical knowledge.

\vspace{3pt}\noindent \textbf{Polar feature extractor module (PFEM):}
To extract shallow features in different views, we propose PFEM, which consists of a multi-kernel atrous convolution module (MKAC), a multi-kernel pooling module (MKPM), and a convolutional block attention module (CBAM) \cite{woo2018cbam}. 
For each projection ${x_i}$, MKAC ${H \left ( \cdot \right ) }$ applies multiple scale atrous convolutions to enlarge the field of view\cite{yang2022recepnet}, 
and MKPM ${G \left ( \cdot \right ) }$ applies a series of max-pooling operations with different pooling kernels to discover microscopic changes, by extracting the most salient feature\cite{ju2022normattention}. 
Finally, CBAM ${T \left ( \cdot \right ) }$ is applied to exploit the inter-channel hidden information of features. 
During feature extraction, ${W_n}$ is used to adaptively adjust the weights of the above processes. 
The mathematical notation of the above is:
\begin{equation}\small
\left\{\begin{array} { l } 
F_{\text {MKAC}}=\operatorname{LeakyReLU}\left [ \sum_{n}H \left( x_{i}\right) W_{n} \right ] ,\\ 
F_{\text {MKPM}}=\operatorname{LeakyReLU}\left [ x_{i} + \sum_{n}G \left( x_{i}\right) W_{n} \right ]  ,\\
F_{\text {PFEM}}= T\left( \operatorname{Concat} \left(F_{\text {MKAC}} , F_{\text {MKPM}} \right) \right ) .
\end{array} \right.  
\end{equation}

\vspace{3pt}\noindent \textbf{Polar region importance module (PRIM):} 
To calculate the region importance, we implement PRIM by applying an average pooling after a Grad-CAM \cite{selvaraju2017grad}. In order to capture the importance of feature map ${k}$ for class ${c}$, we denote ${a_{k}^{c}}$ as the gradient of the score for class ${c}$, with respect to feature map activations ${A^{k}}$ of the last residual layer. The region importance matrix 
$L_{\text {RI}}^{c}$ is given by:
\begin{equation}\small
L_{\text {RI}}^{c}=\operatorname{AvgPool} [ \operatorname{ReLU}  (\sum_{k} \alpha_{k}^{c} A^{k} )  ] .
\end{equation}

\vspace{-10pt}
\section{Experiments and Results}

\noindent \textbf{Data description:}
An in-house dataset was conducted for this study. It includes 199 images from 114 AD patients and 566 images from 291 healthy subjects. All data were collected with the approval of the relevant authorities and the consent of the patients, following the Declaration of Helsinki. All the patients conform to the standards of the National Institute on Aging and Alzheimer’s Association (NIA-AA). The images were captured with a swept-source OCTA (VG200S, SVision Imaging). The images were captured in a 3${\times}$3 mm${^{2}}$ area centered on the fovea. We make sure that the images from a single patient will only be used as training or testing sets once. In the cross-validation experiment subset, we sample the categories from each dataset at the same ratio.

\begin{table}[!t]\scriptsize
\centering
\caption{The detection results (mean$\pm$std) over the in-house dataset.}\label{tabsota}
\resizebox{1\textwidth}{!}{ 
\setlength{\tabcolsep}{3mm}{
\begin{tabular}{lccc} 
\toprule
Model & ACC & AUROC & Kappa\\
\toprule
ResNet-34\cite{he2016deep} & 0.8125$\pm$0.0267 & 0.7960$\pm$0.0479 & 0.4909$\pm$0.0717\\
EfficientNet-B3\cite{tan2019efficientnet} & 0.7942$\pm$0.0104 & 0.7908$\pm$0.0157 & 0.4177$\pm$0.0267\\
ConvNeXt-S\cite{liu2022convnet} & 0.7562${\pm}$0.0138 & 0.5903${\pm}$0.0313 & 0.1660${\pm}$0.0437 \\
${\operatorname{HorNet-S_{GF}}}$\cite{rao2022hornet} & 0.7602${\pm}$0.0113 & 0.5921${\pm}$0.0286 & 0.1738${\pm}$0.0458 \\
VAN-B6\cite{guo2022visual} & 0.7707${\pm}$0.0124 & 0.6911${\pm}$0.0298 & 0.2939${\pm}$0.0485 \\
ViT-Base\cite{kolesnikov2021image} & 0.7904$\pm$0.0183 & 0.7726$\pm$0.0286 & 0.3641$\pm$0.0715\\
SwinV2-T\cite{liu2022swin} & 0.7601${\pm}$0.0117 & 0.7528${\pm}$0.0343 & 0.3242${\pm}$0.0448 \\
MUCO-Net\cite{10.1007/978-3-031-16434-7_66} & 0.7968$\pm$0.0369 & 0.7773$\pm$0.0414 & 0.3985$\pm$0.0789\\
\hline
{\bfseries Polar-Net w/o PFEM } & {\bfseries 0.8191$\pm$0.0140} & {\bfseries 0.8315$\pm$0.0245} & {\bfseries 0.5279$\pm$0.0224}\\
{\bfseries Polar-Net w/o trans} & {\bfseries 0.8388$\pm$0.0142} & {\bfseries 0.8401$\pm$0.0168} & {\bfseries 0.5279$\pm$0.0264}\\
{\bfseries Polar-Net} & {\bfseries 0.8518$\pm$0.0169} & {\bfseries 0.8484$\pm$0.0295} & {\bfseries 0.5766$\pm$0.0685}\\
{\bfseries Polar-Net w prior} & {\bfseries 0.8532$\pm$0.0174} & {\bfseries 0.8523$\pm$0.0320} & {\bfseries 0.5817$\pm$0.0600}\\
\bottomrule
\end{tabular}}}
\vspace{-10pt}
\end{table}

\vspace{3pt}\noindent \textbf{Implementation details:}
We implemented our proposed method with Pytorch. The model was trained on an Ubuntu 20.04 server equipped with two Nvidia RTX 3090 GPUs. We employed Adam as the optimizer with an initial learning rate of 2e-5 and a batch size of 28. We also applied data augmentation by randomly rotating the images by ${\pm}$20 degrees around their centers. The model was trained for 200 epochs. During the transformation, we considered the difference between the left and right eyes and used nearest-neighbor interpolation. The width of the transformed images was resized to 224 pixels. Five-fold cross-validation was employed to fully utilize the data and make the results more reliable. 
Since there is no standard way to convert existing prior knowledge into matrices, for prior knowledge, we manually generated a 4${\times}$2 weight matrix according to the study \cite{xie2023deep}. The weights were 1 by default. The regions with p-values less than 0.05 had a weight of 1.5, and regions with p-values less than 0.01 had a weight of 2. For the entire DVC, the weight was set to 2.

\vspace{3pt}\noindent \textbf{Evaluation and interpretability assessment:}
We evaluate the performance of the model on the test set using the accuracy score (ACC), area under the receiver operating characteristic (AUROC), and kappa. To evaluate the performance, we compared our method with several state-of-the-art methods in the computer vision field and one in the AD detection field. 
Table \ref{tabsota} shows that our method outperforms the others in ACC, AUROC, and Kappa, with an improvement of up to 4.07\%, 5.63\%, and 9.08\%, respectively. 
Prior knowledge did not bring much performance improvement, partly due to the crude method of generating the prior matrices, and partly probably due to the fact that the network's adaptive algorithm may have already learned similar prior knowledge. 
During the testing phase, we activated PRIM and generated a 4${\times}$2 importance matrix for the entire testing set. An inverse operation of the polar transformation was applied to generate the importance map. 

\begin{figure}[!t]
\centering
\includegraphics[width=1\linewidth]{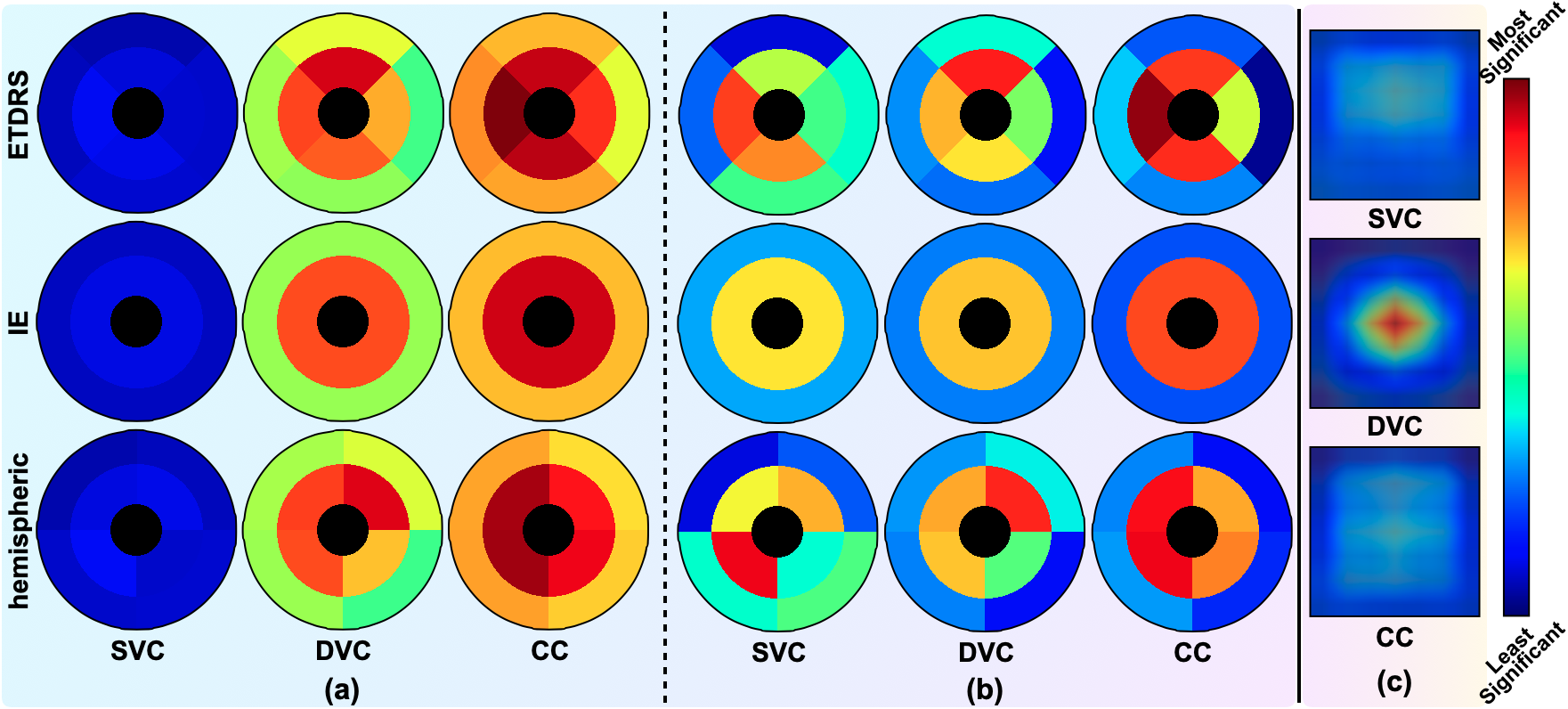}
\vskip -3pt
\caption{Importance maps according to ETDRS, IE, and hemispheric grids. Group (a) takes into account the differences in importance between the projections, while (b) reflects the relative importance of regions within a single projection. The redder the color, the greater the importance. (c) shows a set of examples using the Grad-CAM visualization method, based on ResNet-34 in AD detection.}
\label{figweightcam}
\vspace{-5pt}
\end{figure}

For the sake of simplicity, here we take the EDTRS grid for analysis.
As shown in Fig. \ref{figweightcam} (a), it can be seen that globally, the importance of CC is highest and DVC is the second. This matches the findings that AD patients have a considerably lower density in choriocapillaris flow  \cite{zhang2021choriocapillaris}. 
Meanwhile, the significance of DVC coincides with research findings that there is a considerable reduction in vascular area density and other factors in DVC  \cite{xie2023deep}.
In the DVC and CC, the parafovea is more important. This may relate to the loss of ganglion cells in the parafoveal retina  \cite{un2022posterior}. For single projection (Fig. \ref{figweightcam} (b)), different regions have different importance and the contributions of ${NI}$, ${SI}$ and ${II}$  (illustrated in Fig.\ref{figpolartrans}) are higher. 
In summary, we found a pattern that high importance always occurs where there are more micro-vessels, such as the CC, DVC, and the parafovea. This finding coincides with the conclusion that the  microvasculature of the brain and retina is significantly decreased in AD patients  \cite{zhang2021choriocapillaris}.
Our interpretable results roughly match the clinical study results, because we have made the network follow the clinical analysis method. This also proves the clinical relevance of ours. The minor difference is perhaps because the network unearthed the high-dimensional features that have not yet been discovered clinically.

\begin{table}[!t]\footnotesize
\centering
\caption{The ablation results (mean$\pm$std) of the polar transformation over the in-house dataset.}\label{tabablationpolar}
\resizebox{1\textwidth}{!}{
\setlength{\tabcolsep}{3mm}{
\begin{tabular}{lcccc} 
\toprule
Model & Input & ACC & AUROC & Kappa\\
\toprule
\multirow{2}*{ResNet-34\cite{he2016deep}} & ${w/o \ trans}$ & 0.8125$\pm$0.0267 & 0.7960$\pm$0.0479 & 0.4909$\pm$0.0717\\
~ & $\bm{w \ trans}$ & {\bfseries 0.8244$\pm$0.0226} & {\bfseries 0.8471$\pm$0.0279} & {\bfseries 0.5137$\pm$0.0417}\\
\hline
\multirow{2}*{EfficientNet-B3\cite{tan2019efficientnet}} & ${w/o \ trans}$ & 0.7942$\pm$0.0104 & 0.7908$\pm$0.0157 & 0.4177$\pm$0.0267\\
~ & $\bm{w \ trans}$ & {\bfseries 0.8335$\pm$0.0157} & {\bfseries 0.8295$\pm$0.0279} & {\bfseries 0.5287$\pm$0.0600}\\
\hline
\multirow{2}*{ViT-Base\cite{kolesnikov2021image}} & ${w/o \ trans}$ & 0.7904$\pm$0.0183 & 0.7726$\pm$0.0286 & 0.3641$\pm$0.0715\\
~ & $\bm{w \ trans}$ & {\bfseries 0.7982$\pm$0.0177} & {\bfseries 0.8025$\pm$0.0509} & {\bfseries 0.4235$\pm$0.0748}\\
\hline
\multirow{2}*{MUCO-Net\cite{10.1007/978-3-031-16434-7_66}} & ${w/o \ trans}$ & 0.7968$\pm$0.0369 & 0.7773$\pm$0.0414 & 0.3985$\pm$0.0789\\
~ & $\bm{w \ trans}$ & {\bfseries 0.7916$\pm$0.0255} & {\bfseries 0.7956$\pm$0.0500} & {\bfseries 0.4271$\pm$0.0659}\\
\bottomrule
\end{tabular}}}
\end{table}

\vspace{3pt}\noindent \textbf{Ablation study:}
To evaluate the effectiveness of the polar transformation and Polar-Net, we performed an ablation study. To validate the proposed Polar-Net, we removed the PFEM. The results are shown at the bottom of Table \ref{tabsota}. To validate the transformation, we used the transformed images and the original images respectively. The results are shown in Table \ref{tabablationpolar}. 
All the results showed the effectiveness of the proposed components and modules.

\begin{table}[!t]\scriptsize
\centering
\caption{The classification results (mean$\pm$std) of different methods over the OCTA-500 dataset.}\label{tabocta500}
\resizebox{1\textwidth}{!}{
\setlength{\tabcolsep}{3mm}{
\begin{tabular}{lccc} 
\toprule
Model & ACC & AUROC & Kappa\\
\toprule
ResNet-34\cite{he2016deep} & 0.9641$\pm$0.0389 & 0.9818$\pm$0.0306 & 0.8412$\pm$0.1757 \\
EfficientNet-B3\cite{tan2019efficientnet} & 0.9632$\pm$0.0225 & 0.9741$\pm$0.0245 & 0.8375$\pm$0.1006 \\
VAN-B6\cite{guo2022visual} & 0.9478${\pm}$0.0245 & 0.9517${\pm}$0.0300 & 0.7691${\pm}$0.1334 \\
ViT-Base\cite{kolesnikov2021image} & 0.9692$\pm$0.0281 & 0.9768$\pm$0.0247 & 0.8694$\pm$0.1199 \\
MUCO-Net\cite{10.1007/978-3-031-16434-7_66} & 0.9529$\pm$0.0204 & 0.9717$\pm$0.0222 & 0.8086$\pm$0.0885 \\
\hline
{\bfseries Polar-Net} & {\bfseries 0.9898$\pm$0.0140} & {\bfseries 0.9949$\pm$0.0072} & {\bfseries 0.9604$\pm$0.0544}\\
\bottomrule
\end{tabular}}}
\vspace{-5pt}
\end{table}

\vspace{3pt}\noindent \textbf{Extended experiment:}
To further verify our detection method's stability and generalisability, we conducted an additional experiment on a public dataset OCTA-500 \cite{li2020ipn}.  It contains 189 images from 29 subjects with diabetic retinopathy and 160 healthy control. The details of the implementation are the same as the experiments on the in-house dataset. As shown in Table \ref{tabocta500}, our method achieved the best performances compared to the competitors.

\vspace{-0.2cm}
\section{Conclusion}
\vspace{-0.2cm}
In this paper, we propose a novel framework for AD detection using retinal OCTA images, leveraging clinical prior knowledge and providing interpretable results. Our approach involves polar transformation, allowing for the use of approximate sector convolution and enabling the implementation of the region-based analysis. Additionally, our framework, called Polar-Net, is designed to acquire the importance of the corresponding retinal region, facilitating the understanding of the model’s decision-making process in detecting AD and assessing its conformity to clinical observations.
We evaluate the performance of our method on both private and public datasets, and the results demonstrate that Polar-Net outperforms state-of-the-art methods. Importantly, our approach produces clinically interpretable results, providing a potential tool for disease research to investigate the underlying pathological mechanisms.
Our work presents a promising approach to using OCTA imaging for AD detection. 
Furthermore, we highlight the importance of incorporating clinical knowledge into AI models to improve interpretability and clinical applicability. 

\subsubsection{Acknowledgment.} This work was supported in part by the National Science Foundation Program of China (62272444, 62103398), Zhejiang Provincial Natural Science Foundation of China (LR22F020008), the Youth Innovation Promotion Association CAS (2021298), the A*STAR AME Programmatic Funding Scheme Under Project A20H4b0141, and A*STAR Central Research Fund.

\bibliographystyle{splncs04}
\bibliography{biblib}

\end{document}